\documentclass[ superscriptaddress,bibnotes,amsmath,amssymb, twocolumn, aps]{revtex4-2}

\usepackage{graphicx}
\usepackage{dcolumn}
\usepackage{bm}
\usepackage[usenames,dvipsnames]{xcolor}
\definecolor{darkblue} {rgb}{0.0, 0.0, 0.4}
\definecolor{darkred} {rgb}{0.58, 0.21, 0.20}
\definecolor{darkgreen} {rgb}{0.20, 0.58, 0.21}

\preprint{}

\begin{document}

\title{Field-controlling patterns of sheared ferrofluid droplets}

\author{Shunichi Ishida}
\thanks{S. I. and Y. Y contributed equally to this work.}
\email{Corresponding author. ishida@mech.kobe-u.ac.jp}
\affiliation{Graduate School of Engineering, Kobe University, Kobe 6578501, Japan}

\author{Yaochen Yang}
\thanks{S. I. and Y. Y contributed equally to this work.}
\affiliation{CAS Key Laboratory for Theoretical Physics, Institute of Theoretical Physics, Chinese Academy of Sciences, Beijing 100190, China}
\affiliation{School of Physical Sciences, University of Chinese Academy of Sciences, 19A Yuquan Road, Beijing 100049, China}

\author{Fanlong Meng}
\email{Corresponding author. fanlong.meng@itp.ac.cn}
\affiliation{CAS Key Laboratory for Theoretical Physics, Institute of Theoretical Physics, Chinese Academy of Sciences, Beijing 100190, China}
\affiliation{School of Physical Sciences, University of Chinese Academy of Sciences, 19A Yuquan Road, Beijing 100049, China}

\author{Daiki Matsunaga}
\email{Corresponding author. daiki.matsunaga.es@osaka-u.ac.jp}
\affiliation{Graduate School of Engineering Science, Osaka University, Toyonaka 5608531, Japan}

\date{\today}

\begin{abstract}
We investigate how ferrofluid droplets suspended in a wall-bounded shear flow can organise when subjected to an external magnetic field. 
By tuning the magnitude of the external magnetic field, we find that the ferrofluid droplets form chain-like structures in the flow direction when the magnetic field is weak, while forming a crystal-like pattern in a strong magnetic field. 
We provide the phase diagram and the critical conditions for this chain-to-crystal transition, by applying both numerical simulations and analytic calculations. 
We also examine how the organised patterns of the ferrofluid droplets can be controlled by simply changing the direction of the magnetic field.
This work demonstrates new aspects of field-controllable ferrofluid droplets as a configurable and reprocessable metamaterial.
\end{abstract}

\maketitle

Ferrofluid droplets are magnetic emulsions that are immersed in a non-magnetic fluid, which can be magnetized under a uniform magnetic field and extend in the field direction \cite{Rosensweig1985,Afkhami2010};
such uniform actuation by the magnetic field allows user-friendly applications, which is also widely utilised in designing the collective dynamics of magnetic colloidal \cite{erb2016,ortiz2019,Massana2019,meng2020} and magnetic active systems \cite{Meng2018,Matsunaga2019,Meng2021}.
Differently from typical magnetic colloidal particles, shape of ferrofluid droplets can be controllable by the uniform magnetic field, which are determined by an equilibrium between the surface tension and magnetic force.
As a popular class of magnetic metamaterials, ferrofluid droplets have been utilized for reconfigurable robots \cite{Timonen2013,Wang2018,Fan2020}, controlling the suspension rheologies \cite{Cunha2018,Ishida2020,Abicalil2021}
and cross-streamline migrations \cite{Hassan2018, Zhang2019}. 

\begin{figure}[t]
  \centering
  \includegraphics[width=1\columnwidth]{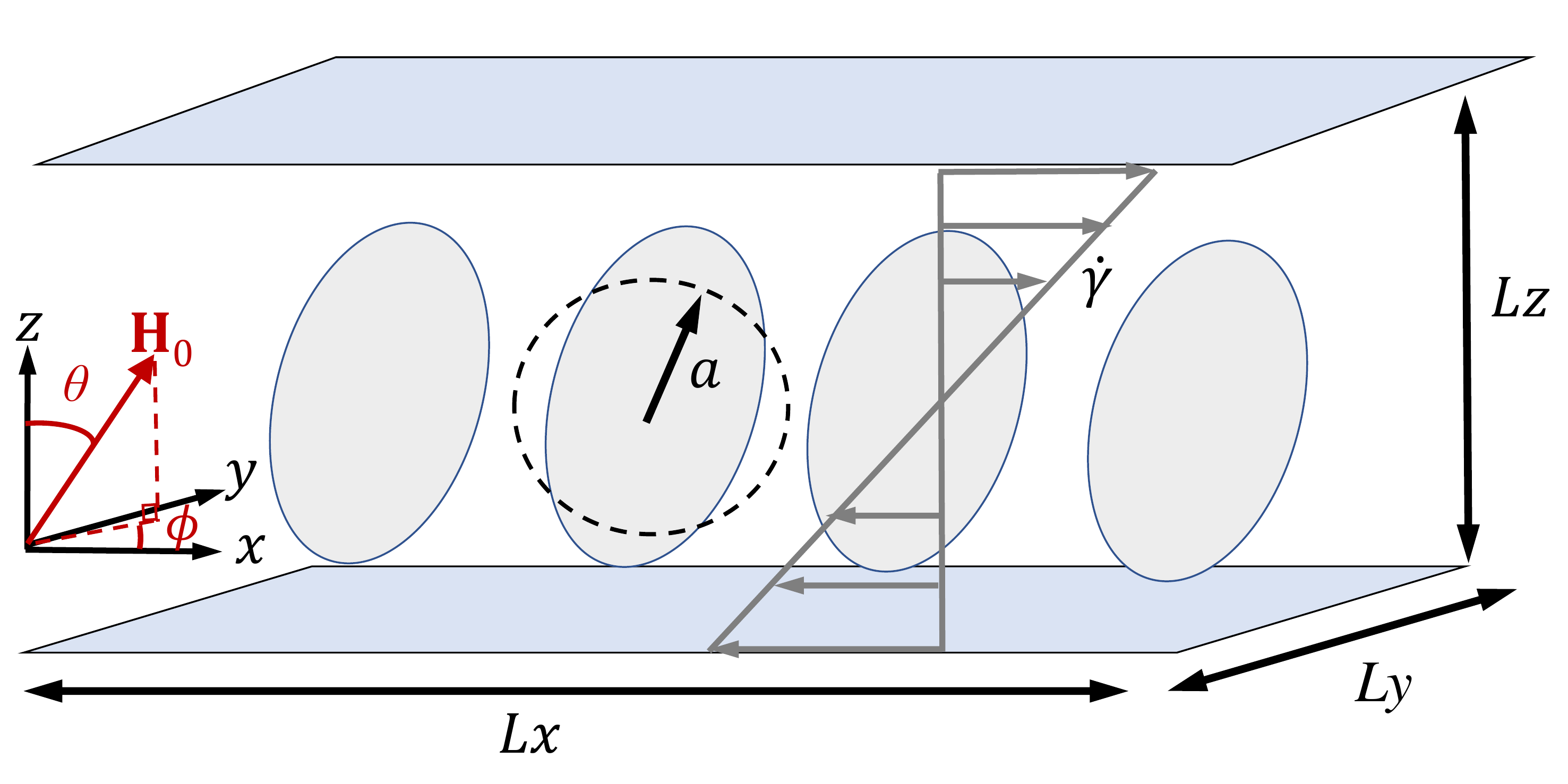}
  \caption{A schematic illustration of the system setup. \label{fig:schematic}}
\end{figure}

A mixture of ferrofluid and non-magnetic fluid under confinements has been known to form various patterns such as labyrinthine \cite{Rosensweig1983} or crystal (hexagonal) pattern \cite{Bacri1988}, which are tunable by magnetic field  \cite{Rosensweig1985,Andelman2009};  
by increasing the strength of the magnetic field, the two-phase fluids can change the organised pattern from the labyrinthine one to the crystal one when the magnetic dipolar interactions dominate over the surface energy \cite{Hong2001,Richardi2002}.
Meanwhile, it has been known that flow field can also help to tune the organisation patterns of suspended non-magnetic particles; for example droplets \cite{Singha2019} or red blood cells \cite{Shen2018} under a wall-bounded shear can form 1D chain or 2D crystal, depending on the area fraction or the wall height. 
Thus, it is not trivial to answer how ferrofluid droplets can be organised when considering both the magnetic interactions and hydrodynamic interactions between the droplets, which is usually the case in microfluidic applications of ferrofluid droplets. 
In this work, we study how ferrofluid droplets suspended in a wall-bounded shear flow can be organised under an external magnetic field, as shown in Fig.~\ref{fig:schematic}. 
By controlling the external magnetic field, e.g., the magnitude and the field direction,
the organised patterns of ferrofluid droplets will be analysed by both numerical simulations and analytical calculations.

\begin{figure*}[t]
  \centering
  \includegraphics[width=1.95\columnwidth]{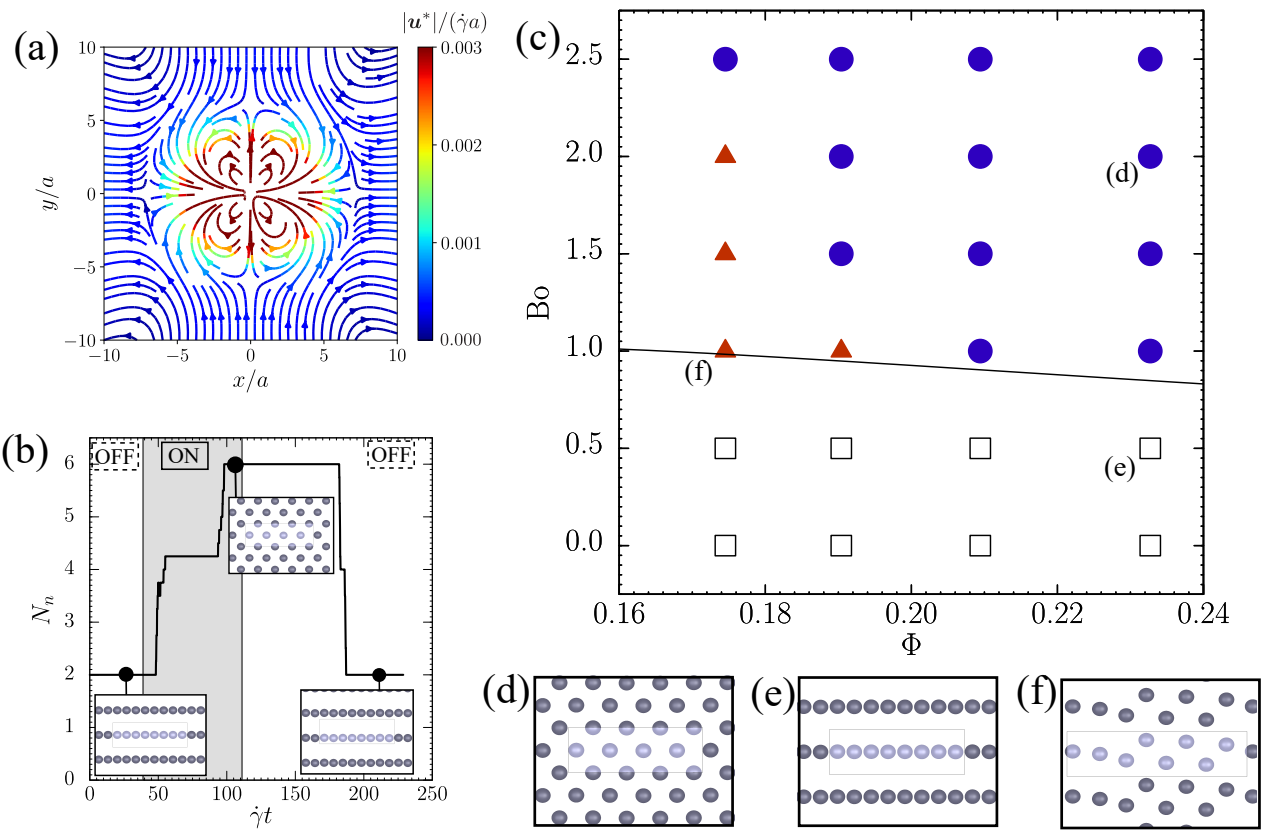}
  \caption{(a) Streamlines at the channel mid-plane due to the dipole force at the origin $\bm{u}^*$ (see supplemental materials). (b) Reversible pattern change in the droplet structure when the magnetic field (+$\mathrm{z}$ direction) is on/off. Note $N_n$ is the number of neighboring droplets (see also \textbf{Movie 1}). In inset simulation figures, gray rectangles show the original computational domain while droplets outside the range are the periodic mirrors. (c) Phase diagram of the spatial patterns when the magnetic field is applied perpendicular to the parallel plates ($\theta = 0$. Conditions are ${\rm Ca}=0.10$, $N = 8$ and $L_\mathrm{y}/a = 6$. Each symbol denotes square (chain pattern), circle (crystal pattern) and triangle (incomplete crystal pattern), respectively. The solid curve is obtained by Equation~\ref{p1} with $c=0.5$. \label{fig:phase}}
\end{figure*}

Consider ferrofluid droplets of radius $a$ suspended between two parallel plates which are separated by a distance $L_\mathrm{z}$ ($L_\mathrm{z} = 3a$ is fixed in the system) as shown in Fig.~\ref{fig:schematic}.
By relative movements of two plates, a simple shear flow with shear rate $\dot{\gamma}$ can be generated.
For simplicity, we assume the viscosity $\eta$ and the density $\rho$ the same inside and outside of the droplets.
The magnetic permeability inside the droplet $\mu_d$ is assumed to be $\mu_d/\mu_0 = 2$ where $\mu_0$ is the vacuum permeability.
An  external magnetic field is applied: $\bm{H}_0 = H_0 (\sin \theta \cos \phi, \sin \theta \sin \phi, \cos \theta)$, and $\bm{H} = \bm{H}_0$ is taken at both the top and the bottom wall as boundary conditions for the magnetic field.
The droplets can be magnetized with the magnetization $\bm{M} = \chi \bm{H}$,
where $\chi$ is the magnetic susceptibility.
Suppose that there are $N$ droplets in the system, then the area fraction of the droplets in \emph{xy} plane is $\Phi = N\pi a^{2}/(L_\mathrm{x} L_\mathrm{y})$, which will be varied from $0.17$ to $0.23$ in this work by changing the domain size $L_\mathrm{x} L_\mathrm{y}$ in our system.
In this system, there are three dimensionless parameters:
Reynolds number ${\rm Re} = \rho \dot{\gamma} a^2/\eta$ comparing the effect of the inertia with that of the viscous stress, capillary number ${\rm Ca} = \eta \dot{\gamma} a/\sigma$ comparing the effect of the viscous stress with that of the surface tension, and Bond number ${\rm Bo} = a \mu_0 H_0^{2}/(2 \sigma)$ comparing the effect of the magnetic energy with that of the surface tension. In the following study, $\rm{Re}=0.1$ and $\rm{Ca}=0.1$ are fixed for simplicity, and how Bond number can change the collective response of ferrofluid droplets is the main focus of this work. 
The two-phase flow hydrodynamics is solved by lattice Boltzmann method \cite{Ishida2020} coupled with the front-tracking method \cite{Tryggvason2001}, and a lattice spacing $\Delta x = a/24$ is adopted.
The periodic boundary conditions are applied in $\mathrm{x}$ and $\mathrm{y}$ directions for the fluid velocities and the magnetic fields.


\emph{Chain-to-crystal transition.}
In a simple shear flow, the droplets will migrate towards the mid-plane of the channel ($z = L_\mathrm{z}/2$) due to the hydrodynamic interaction between the droplets and the walls \cite{Blake1971,Smart1991}.
Since droplets in the wall-bounded shear flow can generate quadrupole flow fields \cite{Shen2018,Singha2019} (see Fig.~\ref{fig:phase}(a)), they will attract and align in the flow direction ($\mathrm{x}$ direction) while they repel each other in the vorticity direction ($\mathrm{y}$ direction).
The droplets have their stable inter-droplet distance due to the interplay with the flow field \cite{Shen2018,Singha2019}.
As a result, the droplets form a chain-like pattern as Ref.~\cite{Migler2001,Singha2019} [see Fig.~\ref{fig:phase}(b)] when there is no magnetic field.
By applying a magnetic field which is perpendicular to the parallel walls ($\theta = 0$; $+\mathrm{z}$ magnetic field), the droplets will be magnetized nearly along the $\mathrm{z}$ direction \cite{Ishida2020,Cunha2020,Abicalil2021}.
By increasing the strength of the magnetic field, the organisation of the droplets can switch from the chain-like pattern to a 2D crystal-like pattern as shown in Fig.~\ref{fig:phase}(b).
This chain-to-crystal transition is reversible (Fig.~\ref{fig:phase}(b));
the average number of the neighboring droplets (droplets with a distance smaller than $d = 1.3 d_0$ where $d_0$ is the average distance of the closest neighbors at each instantaneous time) near a ferrofluid droplet becomes $N_\mathrm{n} = 6$ when the magnetic field is applied, which corresponds to the crystal-like pattern, while the number returns to $2$ once the field is turned off, which corresponds to the chain-like pattern.
The phase diagram of the spatial patterns generated by our numerical simulations is shown in Fig.~\ref{fig:phase}(c).
Note that apart from the chain or crystal patterns, there is an intermediate state where the droplets form incomplete crystal structures as shown in Fig.~\ref{fig:phase}(f), which happens for small area fraction and the magnetic field of intermediate strength.
This pattern emerges since the perfect lattice can not be achieved due to the finite-sized domain in the setup.

\begin{figure*}[t]
  \centering
  \includegraphics[width=2.00\columnwidth]{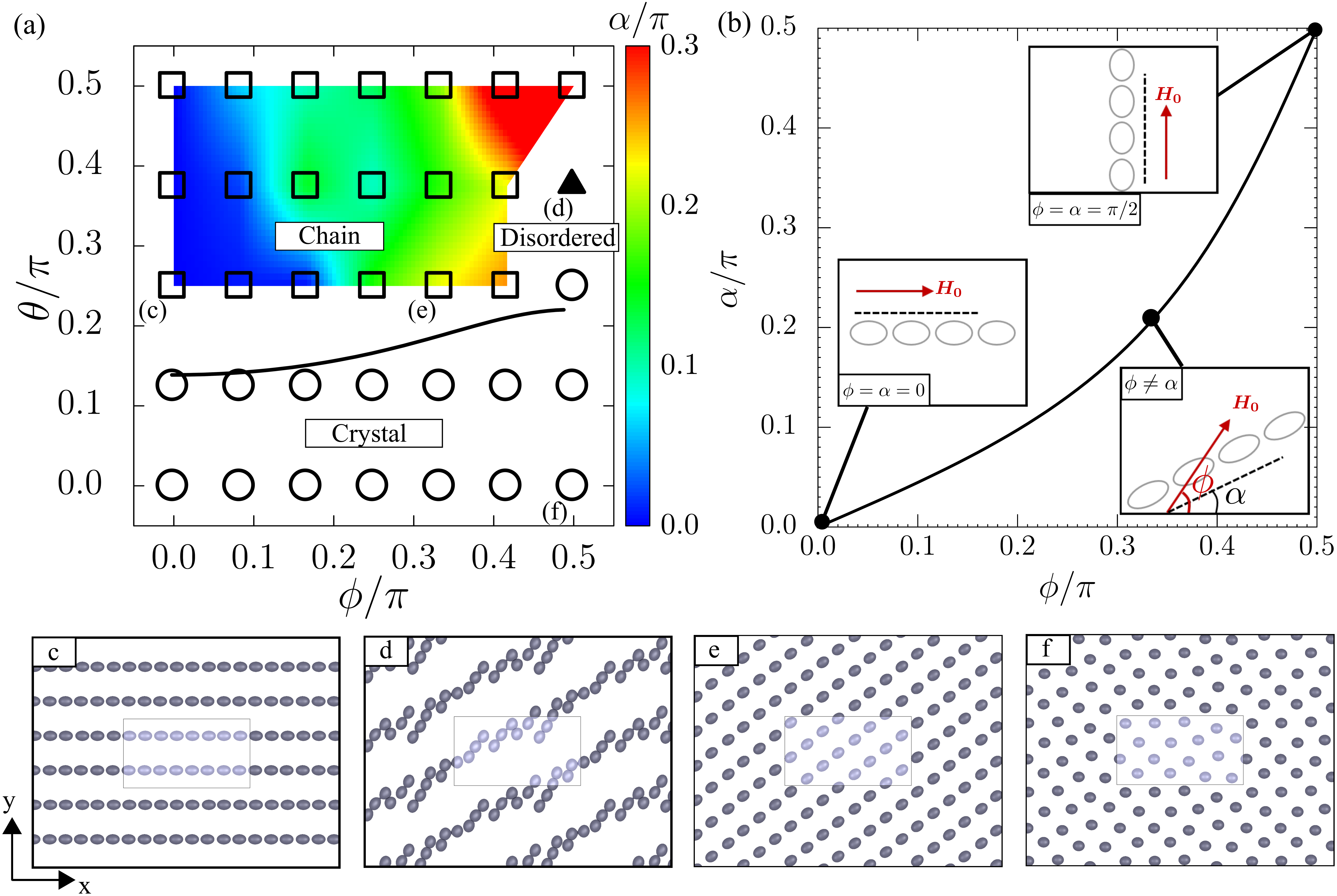}
  \caption{(a) Phase diagram of the spatial pattern under various magnetic field directions under conditions ${\rm Ca}=0.10$, ${\rm Bo} = 2.0$, $\Phi = 0.19$, $N = 16$ and $L_\mathrm{y}/a = 12$. The solid curve is obtained by Equation~\ref{p2} with $c=0.5$. In the phase diagram, each symbol denotes square (chain pattern), circle (crystal pattern) and triangle (disordered pattern), respectively. The contour shows angle of the droplet alignment $\alpha$ in the chain pattern. (b) Relation between $\alpha$ and $\phi$. Figures (c)-(f) are the snapshots from the simulations. \textbf{Movie 2-5} are corresponding movies for Fig.~(c)-(f). \label{fig:diagonal}}
\end{figure*}
%
%


This chain-to-crystal transition can be understood analytically by considering how the hydrodynamic interaction and the magnetic dipole-dipole interaction between the droplets compete.
The quadrupole flow field generated by each sheared droplet can be approximated by the flow field of a force dipole at the mid-plane of two parallel walls, where $F = c \eta \dot{\gamma} \pi a^2$ is the force magnitude with $c$ as a fitting parameter in the model.
Then the flow field of the force dipole is given by $u_j^* = - \bm{\nabla} u_j^{\rm 2W} \cdot \bm{n}$ where $\bm{u}^{\rm 2W}$ is the flow field by a force monopole sandwiched between two walls~\cite{Blake1971} (see supplemental materials for details), $\bm{n} = - 2 a \hat{\bm{x}}$ with $\hat{\bm{x}}$ to be the unit vector in $\mathrm{x}$ direction.
Figure~\ref{fig:phase}(a) shows the flow field generated by the force dipole, and it is clear that the flow $\bm{u}^*$ enters in $\mathrm{x}$ direction and exits in $\mathrm{y}$ direction.
Consider two droplets (\emph{A} and \emph{B}) with the connecting vector $\bm{r}=(\ell_\mathrm{x},0,0)$, and then we can introduce an effective hydrodynamic force $f^\mathrm{h}$ acting on droplet \emph{A} by droplet \emph{B} (note that only x component of this force is non-zero here), as a result of the hydrodynamic interaction between the two droplets (see supplemental materials): 
\begin{equation}
  f_\mathrm{x}^\mathrm{h} = -\zeta\frac{c\dot{\gamma}a^3}{4}\left(\frac{2}{\ell_\mathrm{x}^2}+
\frac{2\ell_\mathrm{x}}{\mathcal{L}^3}-\frac{6\ell_\mathrm{x}^5+9\ell_\mathrm{x}L_\mathrm{z}^4}{\mathcal{L}^7}\right),
\end{equation}
with the friction constant as $\zeta = 6 \pi \eta a$ (droplet assumed spherical), and $\mathcal{L}=\sqrt{\ell_\mathrm{x}^2+L_\mathrm{z}^2}$. 
Note that such hydrodynamic interaction between the droplets is attractive when the two droplets are close to each other. 
When switching on the magnetic field in $\mathrm{z}$ direction, the magnetic force acting on droplet \emph{A} by droplet \emph{B} due to the magnetic dipole-dipole interaction is:
\begin{eqnarray}
  f_\mathrm{x}^\mathrm{m} &=& \frac{3\mu_0 m^2}{4 \pi \ell_\mathrm{x}^4}
\end{eqnarray}
where $m = \pi a^3 H_0$ is the induced magnetic moment of a single spherical droplet by the magnetic field $\bm{H}_0$, and such magnetic dipole-dipole interaction is repulsive.
Then the criterion for the chain-to-crystal transition of ferrofluid droplets can be estimated by solving $f_\mathrm{x}^\mathrm{h} + f_\mathrm{x}^\mathrm{m} = 0$ (balancing the contribution of the two counterparts, hydrodynamic interaction and magnetic dipole-dipole interaction), which reads:
\begin{eqnarray}\label{p1}
\mathrm{Bo} = \mathrm{Ca} \cdot \frac{2 c\pi^2}{\Phi^{2}}\!\Bigg[\frac{\Phi}{\pi}\!+\!\Phi\sqrt{\frac{\pi}{A^3}}\!\left(1-
\frac{6\pi^2+9(L_\mathrm{z}/a)^{4}\Phi^2}{2 A^2}\right)\Bigg]~~~ \nonumber \\
\end{eqnarray}
where $A=\pi+(L_\mathrm{z}/a)^{2}\Phi$ and the characteristic length for the chain-to-crystal transition $\ell_\mathrm{x}^{c}=\sqrt{\pi/\Phi}a$ is taken by assuming that the droplets are arranged in a square lattice at transition.
As shown in Fig.~\ref{fig:phase}(c), the above criterion for the chain-to-crystal transition matches well with the simulation results. 

\emph{Tilted patterns.}
By tilting the magnetic field in terms of the polar and the azimuthal angle, i.e., $\theta$ and $\phi$ as shown in Fig.~\ref{fig:schematic}, $\bm{H}=H_0(\sin\theta\cos\phi,\sin\theta\sin\phi,\cos\theta)$,
the ferrofluid droplets will be magnetised along the magnetic field with the magnitude as $m \approx \pi a^3 H_0$ and the organised patterns will vary.
Figure~\ref{fig:diagonal} shows the phase diagram of the droplet patterns in simulations, with the constant Bond number ${\rm Bo} = 2.0$ and changing values of $\theta$ and $\phi$.
As illustrated in the phase diagram,
the droplets form a crystal pattern when the magnetic field is perpendicular to the parallel plates ($\theta \sim 0$).
When the magnetic field is nearly parallel to the plates ($\theta \sim \pi/2$), on the other hand, the droplets form a chain-like pattern as shown in Fig.~\ref{fig:diagonal}(c) and (e).
Apart from these two limiting cases, the droplet patterns depend on both the polar and the azimuthal angle, $\theta$ and $\phi$.

Following the previous discussions on the case of perpendicular magnetic field, 
we consider two droplets (\emph{A} and \emph{B}) with the connecting vector $\bm{r}=r(\cos\alpha,\sin\alpha,0)$, 
and we can obtain the effective hydrodynamic force acting on droplet \emph{A} by droplet \emph{B}, as:
\begin{eqnarray}
&&\frac{f_\mathrm{x}^\mathrm{h}}{3\pi\eta \dot{\gamma}ac\ell_\mathrm{x}}=\\\nonumber
&&\frac{a^3(1-3\cos^2\alpha)}{2 r^3}+\frac{a^3}{\mathcal{L}^3}\bigg[-1+3\frac{\ell_\mathrm{x}^2}{\mathcal{L}^2}+
\frac{9L_\mathrm{z}^2}{2\mathcal{L}^2}
\left(1-\frac{5\ell_\mathrm{x}^2}{3\mathcal{L}^2}\right)\bigg]\\
&&\frac{f_\mathrm{y}^\mathrm{h}}{3\pi\eta\dot{\gamma}ac\ell_\mathrm{y}}=\nonumber\\\nonumber
&&\frac{a^3(1-3\cos^2\alpha)}{2r^3}+
\frac{a^3}{\mathcal{L}^3}\bigg[-1+3\frac{\ell_\mathrm{x}^2}{\mathcal{L}^2}+
\frac{3L_\mathrm{z}^2}{2\mathcal{L}^2}\left(1-
5\frac{\ell_\mathrm{x}^2}{\mathcal{L}^2}\right)\bigg]
\end{eqnarray}
where $\mathcal{L}=\sqrt{r^2+L_\mathrm{z}^2}$, $r=\sqrt{\ell_\mathrm{x}^2+\ell_\mathrm{y}^2}$, and z-component of the force is not considered as both droplets are focused at the middle plane in z direction. 
Meanwhile, the magnetic dipole-dipole force can be expressed as: 
\begin{eqnarray}
&&f_\mathrm{x}^\mathrm{m}\cdot\frac{4\pi r^4}{3\mu_0m^2}
=\\\nonumber
&&-\sin^2\theta \cos(\phi-\alpha)[5\cos(\phi-\alpha)\cos\alpha-2\cos\phi]+\cos\alpha, \\
&&f_\mathrm{y}^\mathrm{m}\cdot\nonumber\frac{4\pi r^4}{3\mu_0m^2}
=\\\nonumber
&&-\sin^2\theta \cos(\phi-\alpha)[5\cos(\phi-\alpha)\sin\alpha-2\sin\phi]+\sin\alpha.
\end{eqnarray}
By balancing the contributions of hydrodynamic and magnetic dipole-dipole interaction between droplets, 
\begin{eqnarray}\label{p2}
\bm{f}^\mathrm{h}+\bm{f}^\mathrm{m}=0,
\end{eqnarray}
we can obtain the criterion for the chain-crystal transition, 
which is numerically calculated and denoted by the solid curve in Fig.~\ref{fig:diagonal}(a). 
As shown in the figure, ferrofluid droplets can easily form crystal pattern at small tilting angle $\theta$ and form chain-like pattern at large $\theta$. 
Azimuthal angle $\phi$ can also help to tune the organisation of the droplets, which is symmetric regarding $\pi/2$, and the principle direction of the crystal can be changed by the angle $\phi$. 
Note that the value of $\alpha$ is not necessarily equal to $\phi$ as shown in Fig.~\ref{fig:diagonal}(b), i.e., the direction of the formed chains is not the same as that of the magnetic field.

In this work, we analyze organized patterns of ferrofluid droplets suspended in a wall-bounded shear flow.
By applying shear flow and magnetic field, we show that various collective patterns of ferrofluid droplets can be realized by controlling the interplay of hydrodynamic and magnetic interactions.
Without magnetic field, droplets can form a chain-like pattern due to the hydrodynamic interactions between droplets.
If the applied magnetic field perpendicular to the parallel walls is strong enough, ferrofluid droplets can alter their organised patterns from the chain-like to a crystal-like pattern because of the repulsive magnetic interactions between the droplets.
Our system demonstrates the potential of ferrofluids as a configureable and reprocessible metamaterial.

\section*{Acknowledgment}
This work was supported by JSPS (Japan Society for the Promotion of Science) KAKENHI Grants No. 20K14649 and No. 19K20672 and JST (Japan Science and Technology Agency) ACT-X Grant No. JPMJAX190S, PRESTO JPMJPR21OA and Multidisciplinary Research Laboratory System for Future Developments (MIRAI LAB).
A part of the computation was carried out using the computer resource offered
under the category of General Project by Research Institute for
Information Technology, Kyushu University.
This work was also supported by MEXT Promotion of Distinctive Joint Research Center Program Grant Number JPMXP0620335886.
F.M. acknowledges supports from Chinese Academy of Sciences (No. XDA17010504 and No. XDPB15), and the National Natural Science Foundation of China (No. 12047503).

\section*{Movie captions}
\begin{itemize}
  \item {\textbf{Movie 1}: Reversible pattern change in the droplet structure when the magnetic field (+$\mathrm{z}$ direction) is on/off. The magnetic field is on when the droplets are blue while it is off when they are gray. Conditions: $\Phi = 0.23$ and Bo $= 2.0$.}
  \item {\textbf{Movie 2}: Movie of Fig. 3(c). The magnetic field is applied in $\theta = \pi/4$ and $\phi = 0$.}
  \item {\textbf{Movie 3}: Movie of Fig. 3(d). The magnetic field is applied in $\theta = 3\pi/8$ and $\phi = \pi/2$.}
  \item {\textbf{Movie 4}: Movie of Fig. 3(e). The magnetic field is applied in $\theta = \pi/4$ and $\phi = \pi/3$.}
  \item {\textbf{Movie 5}: Movie of Fig. 3(f). The magnetic field is applied in $\theta = 0$ and $\phi = \pi/2$.}
\end{itemize}

\bibliography{reference}

\end{document}